\documentstyle[aps,amssymb,twocolumn]{revtex}

\begin{document}
\author{V. Brouet$^{1},$ H. Alloul$^{1}$, F. Qu\'{e}r\'{e}$^{1}$, G. Baumgartner$%
^{2} $ and L. Forr\'o$^{2}$}
\title{Detection by NMR of a ``local spin-gap'' in quenched CsC$_{60}$}
\address{$^{1}$Physique des solides, UA2 CNRS, Universit\'{e} Paris-Sud, 91 405 Orsay(France) \\
$^{2}$Lab. de Physique des solides semicristallins, IGA, EPFL, 1015 Lausanne (Switzerland)}
\maketitle

\begin{abstract}
We present a $^{13}$C and $^{133}$Cs NMR investigation of the CsC$_{60}$
cubic quenched phase. Previous ESR measurements suggest that this phase is
metallic, but NMR reveals contrasting electronic behavior on the local
scale. The $^{13}$C spin-lattice relaxation time (T$_{1}$) exhibits a
typical metallic behavior down to 50 K, but ndicates that a partial spin-gap
opens for T%
\mbox{$<$}%
50K Unexpectedly, $^{133}$Cs NMR shows that there are two inequivalent Cs
sites. For one of these sites, the NMR shift and (T$_{1}$T$)^{-1}$ follow an
activated law, confirming the existence of a spin-gap. We ascribe this
spin-gap to the occurrence of localized spin-singlets on a small fraction of
the C$_{60}$ molecules.\medskip 
\end{abstract}

The electronic properties of the alkali fullerides A$_{\text{n}}$C$_{60}$
seem contradictory at first sight. A simple independent electron approach
predicts metallic behavior for any n between 1 and 5, because of the
threefold degeneracy of the conduction band (denoted t$_{1u}$ from its
symmetry).\ However, for even n (Na$_{2}$C$_{60}$ and A$_{4}$C$_{60}$)
insulating behavior is observed and may correspond to a band gap associated
with a lifting of the degeneracy of the t$_{1u\text{ }}$level and/or
electronic correlations \cite{Knupfer}. For half-filling (n = 3), a series
of metallic compounds are found, which suggests that the electronic
correlations are not as large as expected because of the narrow electronic
bands. The A$_{3}$C$_{60}$ family exhibit superconductivity which can be
described in the framework of a conventional phonon-mediated mechanism \cite
{Gunarson}.\ From this, one would also expect a simple metallic (perhaps
superconducting) behavior for\ the AC$_{60}$ composition.\ On the contrary,
the cubic high-temperature (CHT) phase of the AC$_{60}$ compounds has a
Curie-like susceptibility \cite{ChauvetPRL94}, suggesting electronic
localization. This is consistent with the nearly T independent NMR
relaxation rates, which indicate further very weak exchange constants
between the localized electrons \cite{TyckoPRB93}. Early optical
conductivity measurements, however, found metallic conductivity \cite
{MartinPRB94}.

The occurrence of a structural transition in AC$_{60}$ from cubic to
orthorhombic symmetry below about 350 K leads to the formation of a
polymerized phase. To extend the study of the CHT phase to lower T, attempts
were made to preserve the cubic symmetry by quenching. For KC$_{60}$ and RbC$%
_{60}$, a phase formed of C$_{60}$ dimers was obtained first \cite{Dimeres}.
More recently, it has been found that a cubic phase (CQ phase for Cubic
Quenched) can be maintained in CsC$_{60}$ \cite{KosakaPRB95} (and RbC$_{60}$ 
\cite{CQRbC60} for faster quenching below 100 K), which transforms
irreversibly into the dimer phase above 130-150 K. In the CHT phase, the C$%
_{60}$ are freely rotating (space group Fm$\overline{3}$m), whereas in the
CQ phase they are static and orientationally ordered (space group Pa$%
\overline{3}$) \cite{LappasJACS95}. Because the ESR spin susceptibility ($%
\chi _{\text{esr}}$) of this phase exhibits no significant T dependence \cite
{KosakaPRB95}, unlike the CHT phase, it was concluded to be metallic. The
inconsistency between the CHT and CQ phases indicates that the properties of
fullerides with one electron per C$_{60}$ are not clear at all, and calls
for more detailed investigations of the CQ phase, which is the only one that
can be studied at low T.

The $^{13}$C and $^{133}$Cs NMR data in the CQ phase presented here not only
evidence a distinct electronic behavior from the CHT phase, but also show
that the results cannot be explained in a simple metallic framework{\it .}\
A partial opening of a spin-gap is detected at low T, which is not
associated with any long range magnetic ordering. Such behavior has not been
observed previously in metallic fullerenes, and thus indicates a new class
of competing electronic instabilities in these compounds. Unexpectedly, we
also find by $^{133}$Cs NMR investigation that there are two inequivalent Cs
sites in the CQ phase. The local electronic susceptibility sensed by one of
these Cs sites is dominated by the existence of the spin-gap, suggesting
that the origin of the two $^{133}$Cs sites is related with the gap.\
\medskip

The CQ phase was obtained by immersing the sample directly into liquid
nitrogen after one hour thermalization at 530~K. NMR measurements have been
performed on the same sample in the polymer, dimer and CHT phases and are
consistent with other results in the literature. ESR and X-ray diffraction 
\cite{Brookhaven} measurements on the sample also confirmed published
results for the CQ phase. No impurity phases could be singled out by either
X-rays or NMR.\ 

An irreversible transition at 135 K between the CQ and dimer phase is
clearly evidenced by $^{13}$C NMR. In the dimer phase, the lineshape is
identical to that of the RbC$_{60}$ dimer \cite{Thierdimere}, i.e. a typical
powder pattern due to anisotropic chemical shifts of sp$_{2}$ carbons. In
the CQ phase, the spectrum is also broad, indicating that molecular motion
is frozen, but it is more symmetric, as observed in metallic fullerides
where an electronic contribution adds to the chemical shift tensor \cite
{Holczer}. This electronic shift is however hard to extract, and the study
of the $^{13}$C relaxation rates (T$_{1}^{-1}$) is more straightforward.

Figure \ref{carT1} shows $^{13}$C (T$_{1}$T)$^{-1}$ in the CQ phase as a
function of temperature. As in other fullerides, the relaxation curves are
not exponential \cite{Holczer}, but they possess the same functional form
over the whole T range (see inset of Figure \ref{carT1}), so that the T
dependence of (T$_{1}$T)$^{-1}$ is unaffected by the choice of the fit
function. In a conventional metal, the NMR relaxation rates usually follow
the Korringa relation (T$_{1}$T)$^{-1}=$ cst. Neglecting the increase of (T$%
_{1}$T)$^{-1}$above 110~K which will be discussed later, (T$_{1}$T)$^{-1}$
is nearly constant between 110~K and 50~K, in agreement with the metallic
character suggested by the Pauli-like $\chi _{\text{esr}}$, and its value
lies in the range of other metallic fullerides \cite{Maniwa}. On the other
hand, (T$_{1}$T)$^{-1}$\ falls by a factor 3 between 50 K and 10 K. Such a
reduction of (T$_{1}$T)$^{-1}$ indicates a loss of weight in the low energy
density of electronic excitations and shows that {\it the CQ phase does not
remain a simple metal down to low T}.{\it \ } \ \medskip

$^{133}$Cs NMR experiments help to clarify the anomalous behavior found by $%
^{13}$C NMR. The spectrum is shown for several temperatures in Fig.\ref
{cesium}. The most prominent feature of these spectra is that {\it there are
two }$^{133}${\it Cs lines}, even though X-ray data demonstrate that there
is a single Cs site in the structure (the octahedral interstitial site of
the fcc lattice) \cite{Brookhaven}. We estimate that the intensities of the
two lines correspond to equal numbers of sites within $10\%$ \cite{quad}.
This ratio is similar in the three other samples investigated. One line
exhibits an unusually large shift of about 800~ppm at 130 K, we denote it
hereafter ``S''\ line for Shifted. The other line is centered at -100~ppm,
which is a typical shift for Cs in fullerides (hereafter ``NS'' line for
Non-Shifted). Although the NS line appears at roughly the same position as
that of the dimer, it can be unambiguously distinguished from the dimer as
it is narrower and its T$_{1}$ is two orders of magnitude smaller at 120~K.
The two lines disappear irreversibly above 135~K, where a single line
characteristic of the dimer phase appears. Thus, the two lines S and NS are
intrinsic to the CQ phase.

To rule out any macroscopic phase separation, we have carried out a double
resonance experiment (SEDOR \cite{sedor,PEnnington}) that allows to detect
sizable dipolar coupling between the two lines. In this experiment, we
observe {\it selectively }the spin-echo of the S line. Normally, the
dephasing caused by dipolar coupling with NS nuclei is refocused and
contributes to the echo intensity. However, when a pulse is applied to NS at
a time $\tau $ after the first pulse on S, a fraction of the S
magnetization, which is determined by the dipolar coupling between this S
site and its NS neighbors, does not refocus. This reduction of the echo
intensity called the sedor fraction (SF) is presented on the inset of fig. 2
for the S line, and follows SF$\varpropto \Delta ^{2}\tau ^{2}$, where $%
\Delta ^{2}$ is determined by the dipolar coupling between S and NS sites.
Since dipolar coupling decays as 1/r$^{3}$, the very fact that we detect a
SEDOR effect implies that the {\it two sites are mixed on the microscopic
scale }\cite{Qsedor}{\it .}

Why do these lines have such different shifts ? Recall that the shift K of
an NMR line is proportional to the local electronic susceptibility ($\chi _{%
\text{loc}}$) through a hyperfine coupling, which is constant for a given
environment and usually T independent. Different shifts could thus arise
from different $\chi _{\text{loc}}$ and/or different hyperfine couplings. In
both cases, the sites must be structurally inequivalent, which is very
puzzling in our case, since a thorough X-ray investigation {\it of the same
sample} showed no deviation from the orientationally ordered Pa$\overline{3}$
cubic structure \cite{Brookhaven}. The situation is somewhat reminiscent of
the A$_{3}$C$_{60}$ case, for which three different alkali NMR lines are
observed at low T, instead of the expected two lines corresponding to filled
octahedral and tetrahedral sites \cite{Walstedt}.

The T dependence of the shifts in the CQ phase is shown in Figure \ref
{shiftalkali}, together with the shifts in other CsC$_{60}$ phases for
comparison. It is immediately clear that neither CQ lines has a T dependence
that could be extrapolated to the Curie law of the CHT phase. Hence, the two
cubic phases have markedly different properties, despite their structural
similarity. However, whereas the shift in the CHT phase scales with the
Curie-like $\chi _{\text{esr}}$, the shift of the $^{133}$Cs S line falls
sharply below 80~K, at variance with the nearly T independent $\chi _{\text{%
esr}}$. This corresponds to a very large T variation of $\chi _{loc\text{ }}$%
at the S site, since the reference chemical shift should be around -200~ppm.
For the NS line, the shift is very small, so that it is rather hard to
determine the T dependence of $\chi _{loc\text{ }}$and to decide whether the
two sites differ only by hyperfine coupling. Relaxation measurements allow
an easier comparison between the different sites.

Figure \ref{allT1} shows (T$_{1}$T)$^{-1}$ on a logarithmic scale for $^{13}$%
C and the two $^{133}$Cs lines. Let us notice first that above 110 K, near
the CQ-dimer transition, all the relaxation rates undergo rapid changes.\
The NS (T$_{1}$T)$^{-1}$ increases by two orders of magnitude, so that the
two $^{133}$Cs lines have similar relaxation rates at 130 K. Note that the
formation of a small fraction of dimer phase would, in contrast, lead to a 
{\it decrease} of the NS\ line relaxation rate. No differences could be
found in X-ray spectra recorded at 25 K and 125 K \cite{Brookhaven}. The
convergence of (T$_{1}$T)$^{-1}$ for the Cs lines should then be ascribed to
an exchange process between the two $^{133}$Cs sites above 110~K. Such an
exchange would produce electronic fluctuations, that could explain the
concomitant increase of (T$_{1}$T)$^{-1}$ for $^{13}$C above 110 K.

Below 110~K, (T$_{1}$T)$^{-1}$ decreases with decreasing temperature for all
the nuclei. The effect is clearest for the S line, and moreover the T
dependence of (T$_{1}$T)$^{-1}$ for this site is identical to that of its
shift, which shows that the static and dynamic properties are similarly
affected. For the NS line, (T$_{1}$T)$^{-1}$ only decreases slightly with
decreasing T, but as T$_{1}$ is already very long at 110~K, it might be
dominated at low T by extrinsic contributions.{\it \ }Even though the $^{13}$%
C and $^{133}$Cs S relaxation rates both decrease with temperature, it is
puzzling to find that they do not scale. For the S line, (T$_{1}$T)$^{-1}$
starts decreasing below 90 K and saturates below 20 K, whereas for $^{13}$C
it begins to decrease below 50~K and does not saturate down to 10K. Although
we are not able to resolve different types of $^{13}$C sites, the splitting
of the $^{133}$Cs spectrum implies the existence of inequivalent C$_{60}$,
so the observed $^{13}$C relaxation is probably an average of two different
behaviors. Better experimental accuracy in $^{13}$C NMR might allow us to
distinguish contributions from different C$_{60}$ balls. Finally, {\it the
combination of the low T decrease of the shift and (T}$_{1}${\it T)}$^{-1}$%
{\it \ of the S line and of the }$^{13}${\it C (T}$_{1}${\it T)}$^{-1}${\it %
\ argue in favor of the existence of a spin-gap}. In the same T range, the $%
^{13}$C and $^{133}$Cs spectra do not broaden, ruling out any magnetic
ordering.

These results imply that the electronic properties of the CQ phase are
microscopocally {\it inhomogeneous}, with coexistence of metallic and
spin-gap behavior. Because the spin-gap behavior is strongest for the S
line, weaker for $^{13}$C, and not presently detected by ESR within
experimental accuracy, we ascribe it to a small fraction of C$_{60}$
molecules, strongly coupled to the neighboring Cs nuclei, which we identify
as S sites. Although only a fraction of the C$_{60}$ are involved in the
spin-gap, let us emphasize that the existence of two very well defined and
non-overlapping $^{133}$Cs lines is very unusual and implies an {\it %
underlying order}. Disorder typically {\it broadens} an NMR line, while the
appearance of two NMR lines implies at least a short range ordering.

The formation of a super-structure seems at first counter-intuitive in this
3D system, in which a strong distortion is ruled out by structural studies.
X-ray investigation of our sample \cite{Brookhaven} shows that the only
possible source of symmetry lowering is the existence of two orientations
corresponding respectively to 85 \% and 15 \% of the C$_{60}$, as in other C$%
_{60}$ Pa$\overline{3}$ compounds. Orientational correlations could be
responsible for a super-structure, with a large unit cell, that could easily
be missed in structural measurements of powder samples. However, if the gap
originates in the band structure of an orientational superlattice, then it
should be sensed equivalently by the different nuclei, and not
preferentially by the Cs S site.

On the other hand, the occurrence of localized singlets strongly
differentiates between the sites depending on their position with respect to
this local perturbation and is in better accordance with our results. Hence,
we use here the denomination of ``local spin-gap''. These singlets could
either be formed in dilute C$_{60}$ dimers or by localization of two
electrons on a ball. In the first case, the dimers would be very weakly
bound as the gap is much smaller than in the quenched dimer phase, for
example. There is no apparent mechanism to stabilize such dimers, and
furthermore, there would be more than two inequivalent Cs sites surrounding
such a structure.

We therefore conclude that the spin-singlets are localized on a C$_{60}$
molecule. The nearest neighbor sites of a C$_{60}^{2-}$ are strongly coupled
to the molecular spin-singlet and constitute the S sites on the observed
spectrum. Since the octahedral Cs site is six-fold coordinated, the
intensity of the S line requires about 10 \% of C$_{60}^{2-}$. Refined ESR
experiments \cite{CQRbC60} are compatible with a decrease by 10 \% of the
ESR susceptibility at low T, which is expected in this model. Such local
electron pairing could be stabilized by a Jahn-Teller (JT) distortion. In
these materials, the energy gain due to a JT distortion is predicted to be
larger for a C$_{60}^{n-}$ with even n, leading to an effective attractive
interaction between electrons for odd n \cite{Victoroff}. Consequently, in
the case of A$_{3}$C$_{60}$, dynamic JT distortions may favor
superconductivity by reducing Coulomb repulsion. However, in CQ CsC$_{60}$,
the two Cs lines indicate that the singlets are static. In this case, an
ordered ground state with alternating C$_{60}^{2-}$ and neutral C$_{60}$
seems at first more natural than the dilution of 10\% C$_{60}^{2-}$ in a
metallic phase. The apparent rigorous stability of this{\it \ dilute singlet
phase }(indicated by the reproducible Cs intensities) may be the result of
frustration in the fcc lattice together with electrostatic repulsion between
C$_{60}^{2-}$, which could make the occurrence of near neighbors C$%
_{60}^{2-} $ energetically unfavorable. Whether these dilute C$_{60}^{2-}$
balls actually form a long-range super-structure remains an open question.
\medskip

In conclusion, we have shown that the CQ phase of CsC$_{60}$ is {\it not} a
simple metal but that a {\it spin-gap} in the electronic structure exists at
low T. We attribute this spin-gap to the formation of localized
spin-singlets on a small fraction of the C$_{60}$, possibly stabilized by a
Jahn-Teller distortion of the C$_{60}^{2-}$ molecules. A structural evidence
for the presence of such JT distorted balls would be welcome, but this may
prove difficult given their low concentration ($\simeq $10 \%). Although the
role of JT distortions has often been invoked to explain the properties of
fullerides - both in the metallic A$_{3}$C$_{60}$ and insulating A$_{4}$C$%
_{60}$ compounds - there is no unambiguous evidence of their importance. Our
results suggest that, in metallic A$_{2\text{p+1}}$C$_{60},$ an insulating
cooperative JT ground state consisting of alternating C$_{60}^{2p-}$ and C$%
_{60}^{(2p+2)-}$ molecules may actually be competing in the extreme case
with a superconducting ground state.

We would like to acknowledge P.W. Stephens and G. Bendele for the X-ray
experiment, M. H\'{e}ritier and A. MacFarlane for helpful discussions.
Financial support from the TMR Programme of the European Comission (Research
Network 'FULPROP' ERBFMRXVT970155) is acknowledged.

\begin{figure}[tbp]
\caption{1/T$_{1}$T measured by standard saturation recovery methods for $%
^{13}$C in the CQ phase. The relaxation curves are integrated over the whole
spectrum and fitted with a stretched exponential (exp(-(t/T$_{1})^\beta$)
with $\beta =0.65$). Inset shows that this fitting procedure holds from 10 K
to 125 K.}
\label{carT1}
\end{figure}

\begin{figure}[tbp]
\caption{$^{133}$Cs spectra in the CQ phase at 10 K, 50 K and 120 K in a 7T
field. Inset : SEDOR fraction at 80 K for the S line as a function of the
time $\tau$ when a $\pi$ pulse is applied to the NS line. The line is a fit
to SF = a+$\Delta ^{2}\tau ^{2}$.}
\label{cesium}
\end{figure}

\begin{figure}[tbp]
\caption{Shifts of the $^{133}$Cs lines with respect to that of a diluted
CsCl solution in the different phases of CsC$_{60}$. The dotted line
extrapolates the Curie behavior of the CHT phase.}
\label{shiftalkali}
\end{figure}

\begin{figure}[tbp]
\caption{Comparison of 1/T$_{1}$T on a logarithmic scale for $^{13}$C and
the two $^{133}$Cs lines in the CQ phase. For $^{133}$Cs, the relaxation
curves are exponential even at low temperatures when they begin to merge
together.}
\label{allT1}
\end{figure}


\begin{references}
\bibitem{quad}  We found that the pulse lengths optimizing the echo
intensity were similar for the two lines, and indicative of small first
order quadrupole effects, which do not distort significantly the measured
intensities.
\end{references}
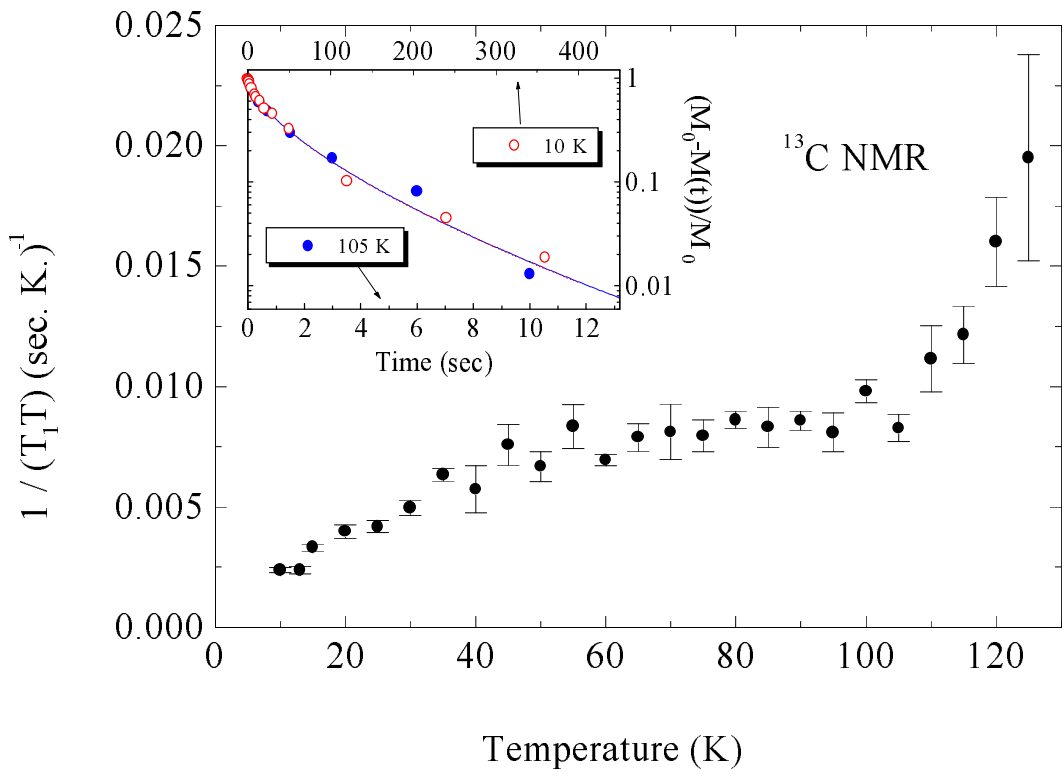
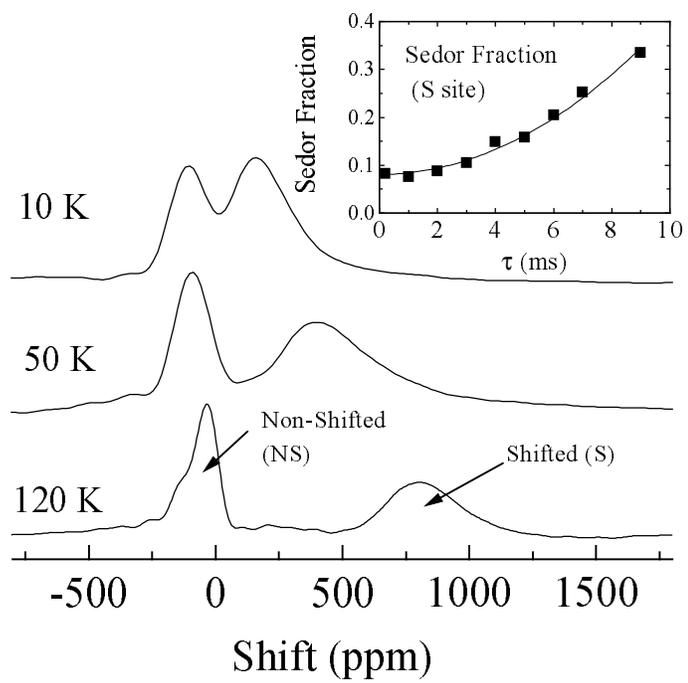
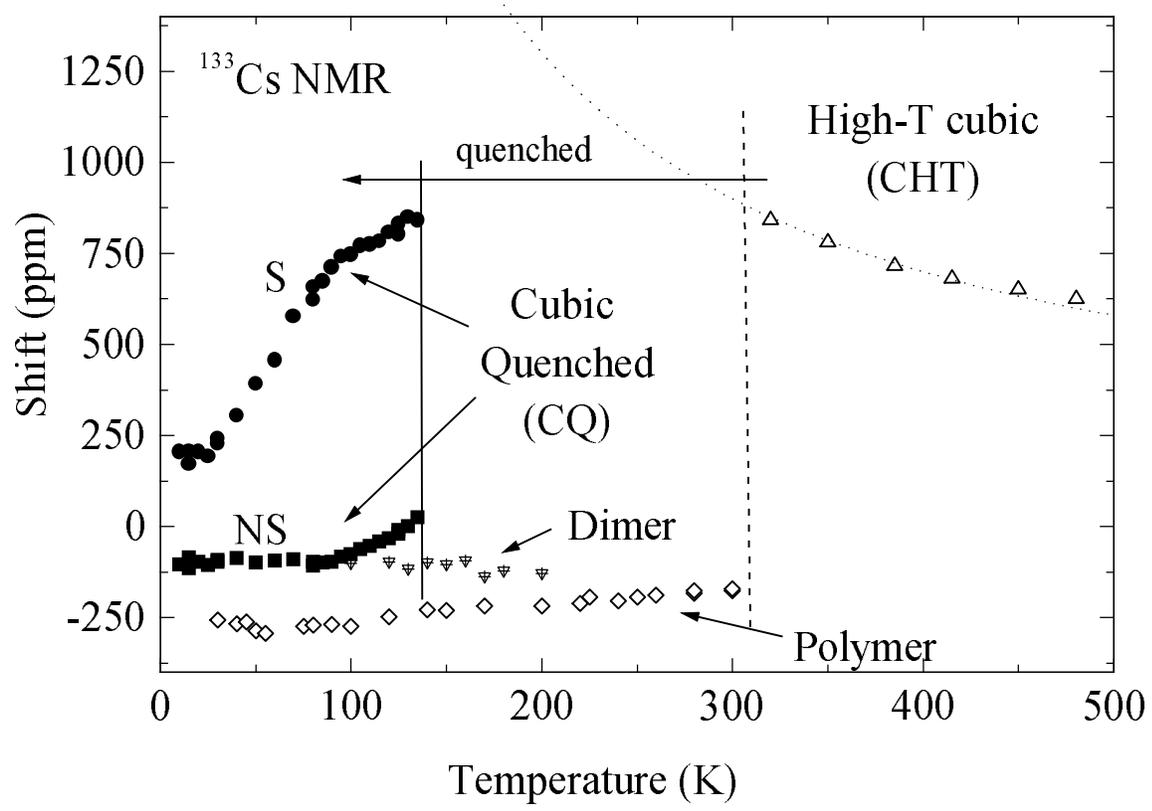
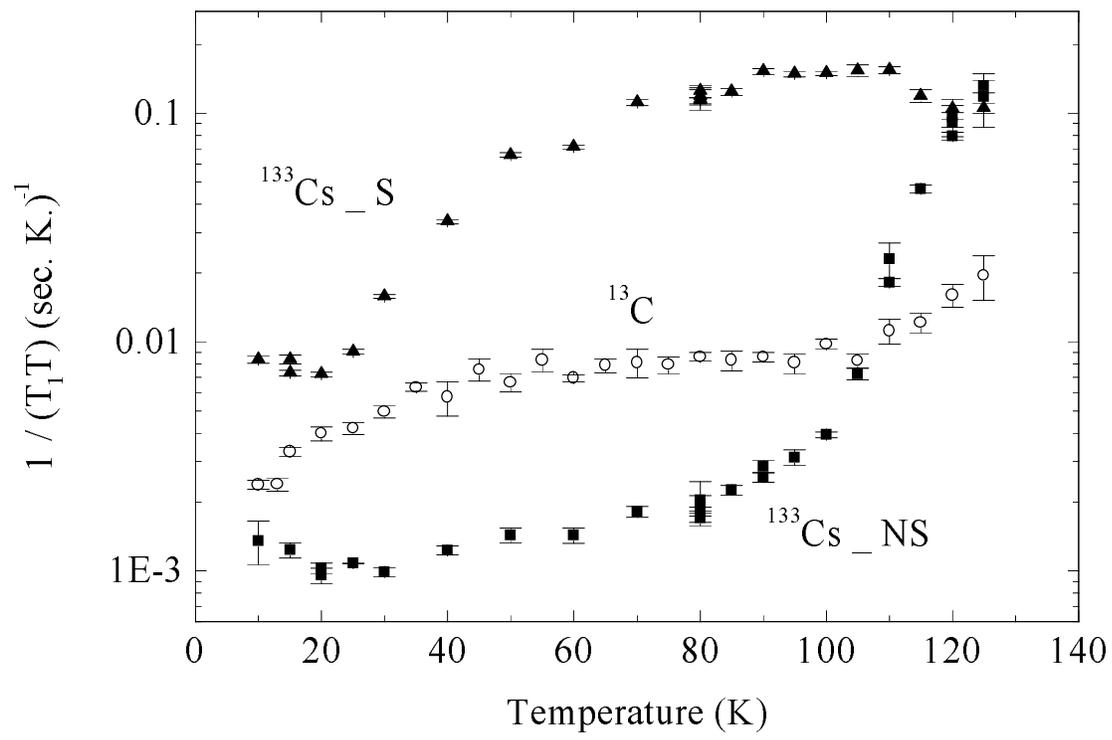
\end{document}